# around Zwicky clusters

Stella Seitz & Peter Schneider
Max-Planck-Institut für Astrophysik
Karl-Schwarzschild-Str. 1
Postfach 1523
D–85740 Garching
Federal Republic of Germany

## Abstract

In a recent paper by Rodrigues-Williams & Hogan (RH94), a correlation between high-redshift, optically-selected QSOs and Zwicky clusters was reported at a very high significance level. Due to the fairly bright flux threshold of the cluster sample, these correlations cannot be interpreted as being due to an environmental effect of the clusters on the quasar activity. The most likely interpretation employed in RH94 was the effect of gravitational lensing by the foreground clusters, though the required magnification to explain the observed correlations has to be considerably higher than obtained from simple mass models for the clusters. We have repeated the analysis of RH94 using a different QSO sample, namely radio quasars and radio galaxies from the 1-Jy sample. In accordance with RH, we detect a statistically significant correlation between Zwicky clusters and 1-Jy sources with intermediate redshift ($z \approx 1$), but fail to detect significant effects for higher-redshift sources. In addition, we detect a highly significant underdensity of low-redshift radio sources around Zwicky clusters, for which an environmental interpretation seems to be most reasonable. Our result for the overdensity of $z \sim 1$ sources is in good agreement with previous results and can possibly be interpreted as a lensing effect, though we have not tried to quantitatively investigate a lensing model. Other interpretations, such as patchy dust obscuration, can not explain the observed effect and its tendency with redshift.

## 1 Introduction

The association of the angular positions of a set of foreground objects with a set of background objects is seen as evidence for the occurrence of a magnification of the background sources due to the (matter associated with the) foreground objects. In fact, provided that the separation along the line of sight of foreground/background objects is larger than the largest coherent cosmological structures, and that the interpretation of redshift as a distance measure is adopted (for an alternative view, see Arp 1987 and refences therein), the magnification effect is the most plausible physical explanation for such an association.

On the scale of a few arcseconds, an overdensity of high-redshift QSOs around low-redshift galaxies has been reported by several groups (e.g., Tyson 1986, Fugmann 1988 –

Wu 1994). It is fair to say that the current observational status it not well understood. On the theoretical side, several studies (Canizares 1981, Vietri & Ostriker 1983, Schneider 1989, Narayan 1989, Kovner 1989, 1991; for a recent summary, see Narayan & Wallington 1994) have shown that one expects an overdensity of galaxies around high-redshift QSOs from lensing by the galaxies, but that the magnitude of this effect depends strongly on the selection criteria of both classes of objects, in particular on the apparent magnitude of the QSOs, and that the expected strength of the effect is smaller than the observational numbers quoted by some of the groups.

On a much larger scale, an association of high-redshift quasars with foreground galaxies has been claimed by Fugmann (1990) on a scale of $\sim 30$ arcminutes. In a series of papers (Bartelmann & Schneider 1993a,b, 1994, Bartelmann, Schneider & Hasinger 1994) it was investigated whether these associations can be due to lensing by large-scale matter inhomogeneities in the universe with which the galaxies are associated (i.e., matter overdensities on scales at least as large as clusters of galaxies; lensing by the individual galaxies can be neglected due to the large angular separations), and it was verified that there are statistically significant associations of high-redshift radio quasars with Lick galaxies, IRAS galaxies, and X-ray photons taken from the ROSAT All Sky Survey. An analytical investigation of the expected effect has recently been completed by Bartelmann (1994) where further applications of these associations were pointed out.

On about the same angular scale, Rodrigues-Williams & Hogan (1994; hereafter RH94) have reported on a statistically significant overdensity of high-redshift QSOs around Zwicky clusters. The overdensity claimed amounts to a factor of about 1.7 and differs from unity at a formal $4.7\sigma$ level. Such a high overdensity is much larger than expected from the lensing action of clusters, as noted by RH94; in addition, this result is in apparent conflict with previous results (Boyle, Fong & Shanks 1988) which indicate an anticorrelation of high-redshift UVX QSOs with cluster galaxies. Furthermore, RH94 briefly noted an underdensity of low-redshift QSOs around Zwicky clusters.

These observational facts are difficult to understand solely within the lensing picture, and RH94 mention several alternative interpretations. Foreground obscuration (e.g., by dust in our galaxy) could account qualitatively for the positive correlations, whereas obscuration in the clusters could yield a local underdensity of QSOs. However, to combine the different pieces of observational evidence into a single explanation scheme is at least a difficult task.

A completely different point of view can be taken by investigating the over- or underdensity of low-redshift AGNi around galaxies and/or clusters for which the redshift is not determined. In that case, the associated galaxies are assumed to be at the same redshift as the AGN and are thus physically associated (e.g., Yee & Green 1984, 1987). This approach then yields information on the effect of the environment of AGNi on their nuclear activity. However, without redshift measurement of the associated galaxies it remains unclear whether they are at the same redshift as the AGN, or in the foreground. This uncertainty particularly prevails for AGNi of intermediate redshift. For example, the overdensity of galaxies around QSOs with redshift $z \sim 1$ detected by Tyson (1986) was interpreted by him as a sign of a strong luminosity evolution of the galaxies which he assumed to be at the same redshift as the QSOs.

In this paper we want to reconsider the associations of high-redshift QSOs with Zwicky clusters, to see whether we can verify the results of RH94. For this purpose, 1981). Nearly all optically identified sources in this catalog with redshift $z \geq 0.5$ are flat-spectrum radio QSOs. The reasons for redoing the analysis with this new set of sources are multiple: (1) radio-selected QSOs are less affected by obscuration; their selection is unaffected by dust. Although the optical identification still may be affected by dust, we nevertheless expect the 1-Jy radio sample to be a much better sample for the present purpose than the optically-identified QSOs used by RH94. (2) If we impose an optical flux threshold onto the radio QSOs, in addition to the radio flux threshold, the multiple magnification bias (Borgeest, von Linde & Refsdal 1991) can cause a much larger magnification bias than for a source sample selected by a flux threshold in a single waveband. We have attributed the observed correlation of 1-Jy QSOs with Lick- and IRAS galaxies to this effective steepening of the source counts. (3) The 1-Jy radio QSOs have already shown signs of a double magnification bias in our previous studies (Bartelmann & Schneider 1993b, 1994; Bartelmann et al. 1994, see especially Fig. 9 in this reference). Taking the latter two points together, we expect that if lensing causes a significant overdensity of Zwicky clusters around high-redshift QSOs, it should most clearly show up for the 1-Jy sources. (4) From Fig. 4 and Table 1 of RH94 it can easily be seen that the distribution of Zwicky clusters in the fields used in the study are distributed highly non-randomly. In fact, from the mean number density of Zwicky clusters on the sky one can see that only two of the nine fields studied by RH94 have a cluster density close to the mean, whereas the other seven fields are highly underdense. (5) The Zwicky catalog of clusters becomes less complete as one approaches the ecliptic equator and the northern ecliptic pole (see Fig. 2a below). All fields studied by RH94 lie close to the ecliptic equator and they should therefore be affected by this incompleteness. Taken together, these last two points imply that the statistical fluctuations are expected to decrease substantially if an all-sky QSO sample is used instead of a QSO sample in some selected regions. (6) Some of the statistical considerations in RH94 are questionable, as we shall discuss in Sect. 4, and their stated significance level is too high an estimate.

We find an overdensity of 1-Jy QSOs with a redshift of order unity around Zwicky clusters at an approx. 98% level and we find no significant effect for higher-redshift QSOs, and a significant (up to 99.89%) underdensity for 1-Jy sources with redshift below 0.5. The significance levels are obtained by extensive Monte-Carlo simulations which account for the non-uniformity of the distribution of the Zwicky clusters.

The rest of this paper is organized as follows: we describe the selection of our quasar and cluster samples in Sect. 2, and the method used for our statistical analysis in Sect. 3. We present the results in Sect. 4, and interprete and discuss them in Sects. 5 and 6, respectively.

## 2 Cluster and Quasar samples

The background sources in our study were taken from the 1-Jansky catalog (Kühr et al., 1981), which was communicated to us in the version of May 1992 by M. Stickel and supplemented by additional optical identifications obtained by Stickel & Kühr (1993a,b). This catalog contains all radio sources brighter than 1 Jansky at 5 GHz which do lie outside galactic plane, $|b| \geq 10°$, and not and not towards the Magellanic Clouds. 424 of these sources are optically identified and have spectroscopically determined redshifts. Most source sample is dominated by radio galaxies. In the following we are not interested in the nature of the radio sources, thus we refer to all of them as 'quasars'. The quasar sample investigated here contains all the 236 optically identified sources in the catalog with determined redshift which have a nonnegative declination, $\delta \geq 0°$; the selection area of these quasars ($|b| \geq 10°$, $\delta \geq 0°$) then covers about 40 percent of the sky. Several subsamples in redshift space were investigated with either no optical flux limit, or a flux limit of 19$^{th}$ and 18$^{th}$ magnitude in the V-band, see Fig. 1.

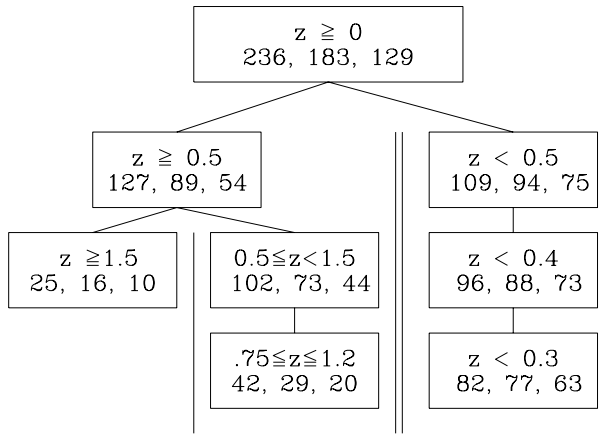

**Fig. 1.** The quasar samples investigated: each box represents three samples containing quasars within the indicated redshift range; one sample consists of quasars where no explicit optical flux limit is imposed, the remaining two contain quasars brighter than 19-th and 18$^{th}$ magnitude, respectively. The three numbers in the second row of each box denote the number of quasars for these three cases

As in RH94, the clusters of galaxies were taken from the Zwicky (1961-68) Catalog of Galaxies and Clusters of Galaxies; they have been selected in those regions of the sky with $\delta \geq -3°$, $|b| \geq 10°$; hence their selection area slightly exceeds that of the quasar sample. In order to remove very nearby clusters, we restrict the cluster sample to those clusters with a Zwicky radius less than or equal to 15 arcminutes. This yields 6342 clusters with a mean value for the Zwicky radius of 8.5 arcminutes, and a distribution in distance classes shifted to Very Distant and Extremly Distant clusters: (4 Near, 100 Medium Distant, 507 Distant, 1928 Very Distant and 3803 Extremely Distant Clusters); in agreement with RH94 we estimate the mean redshift of the cluster sample to be 0.2. For details concerning the correspondence of Zwicky clusters and Abell clusters, distance class-redshift relation or redshift-Zwicky radius relation, see RH94, Sect 2 and 3, and references therein.

Zwicky clusters are not uniformly distributed over their selection area $\delta > -3°$ and $|b| > 10°$. This is illustrated in Fig. 2a, where we have plotted the number of Zwicky clusters contained in a range of declination of width 1°, devided by $\cos\delta$, as a function of declination $\delta$. In addition, we have randomly distributed 6342 points within the defining area of the Zwicky catalog and plotted their corresponding $\delta$-distribution also in Fig. 2a. If the galactic plane were not cut out, a random distribution would then correspond to a horizontal line. The depletion relative to that (fictitious) line for $\delta \lesssim 70°$ is caused by the galactic band and it is strongest at its turnover, i.e. in a range of $50° \lesssim \delta \lesssim 70°$; see the solid curve in Fig. 2a which was obtained by ten time as many points and a rescaling by 0.1. The distribution of the Zwicky clusters in declination does not follow that of a random point distribution : i) the variations of the dotted curve with $\delta$ are much larger than those obtained from a random distribution of the same number of points, see dashed dotted curve. ii)There is a lack of clusters for $\delta \lesssim 10°$ and $\delta \gtrsim 70°$. The first item may be explained by variations in the sensitivity of plates or eyes; an additional source of variation is the correlation between Zwicky clusters. The decrease of the number density of clusters towards high and low values of $\delta$ can be attributed to atmospheric extinction, since the zenith of the Palomar Observatory is at declination $\delta \approx 33°$, and hence the catalog is expected to be most complete at $20° \lesssim \delta \lesssim 40°$, and to show a depletion for the remaining declinations. Hence, we estimate (from the Fig. 2) that the latter effect causes the Zwicky catalog to be incomplete by about 15 percent. From this consideration and Fig. 2a it is evident that selecting $\delta \approx 0°$ fields for studying the density of QSOs near to the line of sight to Zwicky clusters yields very untypical cluster fields; e.g. from the nine fields investigated by RH94 at $\delta \approx 0°$ two fields contain the expected number of clusters whereas the remaining seven fields are depleted by about 50 percent.

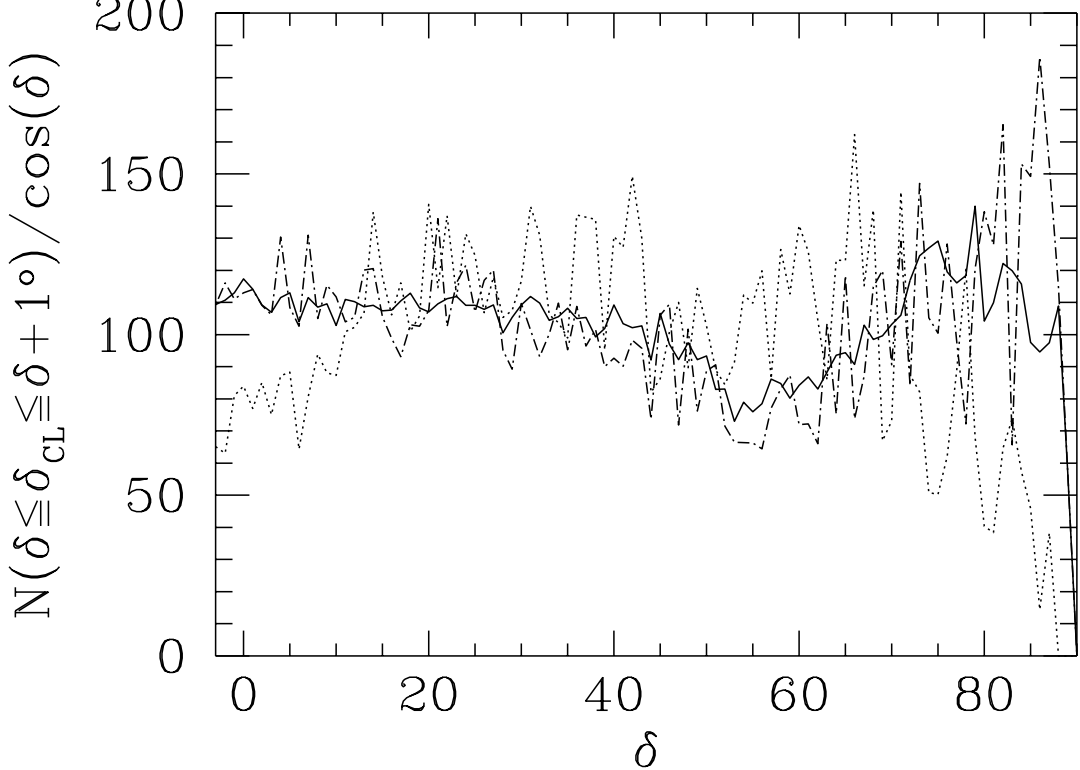

**Fig. 2a.** Number of clusters (dashed curve) with $\delta \leq \delta_{\rm cl} \leq \delta + 1°$ for $-3° \leq \delta \leq 89°$, divided by $\cos\delta$. The same function for 6342 randomly distributed points (dashed-dotted curve) with the same selection criteria as the clusters and with ten time as many points and a rescaling by 0.1 (solid curve)

## 3 Statistical method

objects, one can either relate the density of foreground objects near quasar positions to their average density, or one can relate the density of quasars around cluster positions to the average quasar density. Since quasars are rare, the first strategy is more reasonable and thus usually chosen; however, one then often has to fight with the fact that galaxy densities are not accurately known and vary from plate to plate. To decide whether there is a density enhancement or reduction of background objects close to the line of sight to foreground objects, different methods have been used:

1) For large angular scales (some 10 arcminutes) and independent of the QSOs sample size: The analysis of counts of foreground objects in 25 quadratic or ringshaped cells (the central cell is centered on the background object) with a rank-order statistic (see Fugmann 1990, for an improved statistical method, see Bartelmann & Schneider 1992). Since this 'local' method is only sensitive to a density gradient of the foreground distribution, it does not require the mean density of the foreground population to be known. One can quantify the hypothesis of 'no correlation' with a well-defined error level.

2) For small angular scales (10 to 30 arcseconds) and large QSO samples: One can calculate the density of foreground objects in a circle (Crampton et al. 1992, Yee 1992, Magain et al. 1992), or in a ring (Webster et al. 1992, van Drom 1992) centered on the background object and compare the result with the mean density of the foreground objects considered. This often results in comparing two uncertain quantities with each other.

3) For large QSO samples and large angular scales: One can study the two-point cross-correlation function of background objects with foreground objects (RH94, Romani & Maoz 1992, Boyle et al. 1988, Boyle & Couch 1993). To obtain this function, one compares the number of foreground objects within angular rings centered on QSOs and with those centred on random points. One can take into account the clustering of the quasars by choosing 'random' points which have the same clustering length as the QSOs.

4) For large QSO samples: The nearest-neighbor distance of background QSOs to foreground objects is calculated and compared with the result obtained if the QSOs are replaced by a randomly distributed population (Drinkwater et al. 1991) or, if quasar clustering is important, by 'random' points with the same clustering properties.

5) For large QSO samples and large angular scales a new method was introduced by RH94 to investigate QSO densities relative to foreground Zwicky clusters:

The basic idea is that background sources which are not correlated with foreground objects must be randomly distributed with respect to the foreground objects. If we define an area of the sky in terms of the positions of the foreground objects, the number density of background objects in this area is expected to be the same as the average number density of background objects in the absence of correlations between foreground and background sources. A physically motivated definition of such an area is given by a disc centered on each cluster with an angular radius proportional to the Zwicky radius $\theta_{\rm ZW}$ of the cluster. This ensures that for similar clusters at different distances one examines approximately the same physical scale. One then compares the density of QSOs in this 'cluster area' or 'association area' with that in the remaining area (RH94) or the average quasar density (this paper). The errors in the analysis can be kept small if one takes QSOs from a complete QSO catalog with a sufficient number of objects; an incompleteness in the catalog of the foreground population does not enter the analysis.

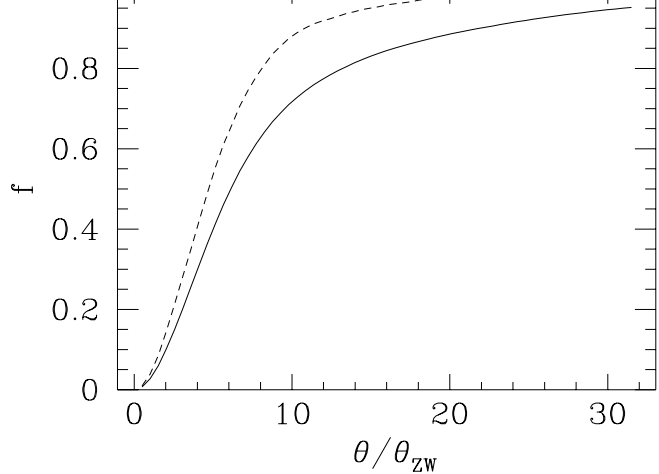

**Fig. 2b.** The fraction $f$ of the total selection area covered by the association area for clusters with $\delta \geq -3°$ and $|b| \geq 10°$ (solid line) and for clusters with $20° \leq \delta \leq 30°$ and $|b| \geq 20°$ (dashed line)

However, if many clusters are missing, this complicates a quantitative interpretation (absorption by dust, gravitational lensing) of the results of the statistical analysis.

We have calculated (with Monte-Carlo integration) the fractional association area $f(x = \theta/\theta_{\mathrm{ZW}})$, i.e., the ratio of the area covered by circles of angular radius $x\theta_{\mathrm{ZW}}$ centered on each Zwicky cluster to the total selection area of the quasars, as a function of $x$, and plotted the result as the solid curve shown in Fig. 2b. For small scales, it increases quadratically since the individual cluster discs do not overlap; at $\theta/\theta_{\mathrm{ZW}} = 3$, 6 and 10, approximately 20, 50 and 70 percent, respectively, of the total area are covered by the association area. The saturation to 100 percent coverage is reached very slowly, since the Zwicky catalog is very incomplete at low galactic latitudes and moderately incomplete at very low and very high declinations; this fact, and also the clustering of clusters causes the distribution of Zwicky clusters on the sky to be highly non-Poissonian. For clusters in the region $20° \leq \delta \leq 40°$ and $|b| \geq 20°$ this improves; they almost completely cover the total field at a scale of $\theta = 20\theta_{\mathrm{ZW}}$ (dashed curve in Fig. 2b).

## 4 Results

Let $N$ be the number of quasars in a particular subsample, $\Omega$ the solid angle of the sky in which the 1-Jy quasars are selected, and $N_{\mathrm{a}}(f)$ the number of quasars within the association area. Then the density of quasars is $n = N/\Omega$, and the densities of associated and non-associated quasars are, respectively

$$n_{\mathrm{a}} = \frac{N_{\mathrm{a}}}{f\Omega} \quad , \quad \bar{n}_{\mathrm{a}} = \frac{N - N_{\mathrm{a}}}{(1-f)\Omega} \quad , \tag{1}$$

respectively. We define the *overdensity* $\omega$ to be

$$\omega := \frac{n_{\mathrm{a}}}{\bar{n}_{\mathrm{a}}} \quad , \tag{2}$$

and the *density enhancement* $q$ as

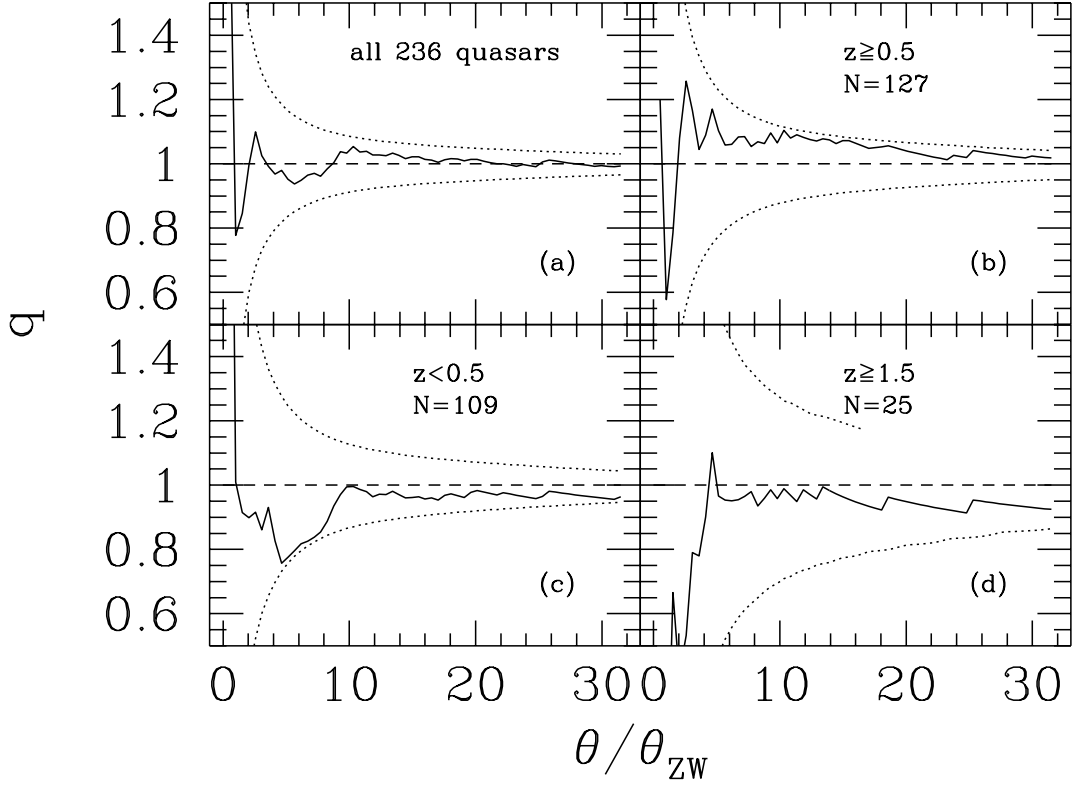

**Fig. 3.** The density enhancement $q$ (solid curve) is plotted as a function of $x = \theta/\theta_{ZW}$ for several quasar samples: (a): all 236 quasars; (b): all quasars with $z \geq 0.5$ (127 quasars); (c): all quasars with $z \leq 0.5$ (109 quasars); (d): all quasars with $z \geq 1.5$ (25 quasars). In all subsamples shown, no optical flux threshold was applied. The dotted curves in each panel correspond roughly to a 98% significance level for a density enhancement or depletion on a given angular scale (or better, given value of $x$). The dashed curve shows the expectation value of $q$ for a random distribution of points. For further details, see the text

Contrary to many authors we do not evaluate the *overdensity* $\omega$ which describes the ratio of the QSO density $n_a$ in the association area and the density in the remaining area. Instead we use the *density enhancement* $q$, i.e., the ratio of the QSO density on the association area to the mean QSO density.

The overdensity is a useful quantity for *interpreting* deviations of the QSO distribution from random by the action of the intervening mass distribution: then, $\bar{n}_a$ is assumed to be the *true* QSO background density, and the observed density near to the line of sight to clusters, $n_a$, has to be deduced from the true counts taking into account observational biases, e.g., dust in clusters or gravitational light deflection. However, since the number of quasars in the non-associated area is also subject to statistical fluctuations, the density obtained will in general not correspond to the true background quasar density.

For the *statistical analysis* one has to evaluate the probability that the observed density $n_a$ in the association area is compatible with a random distribution of background

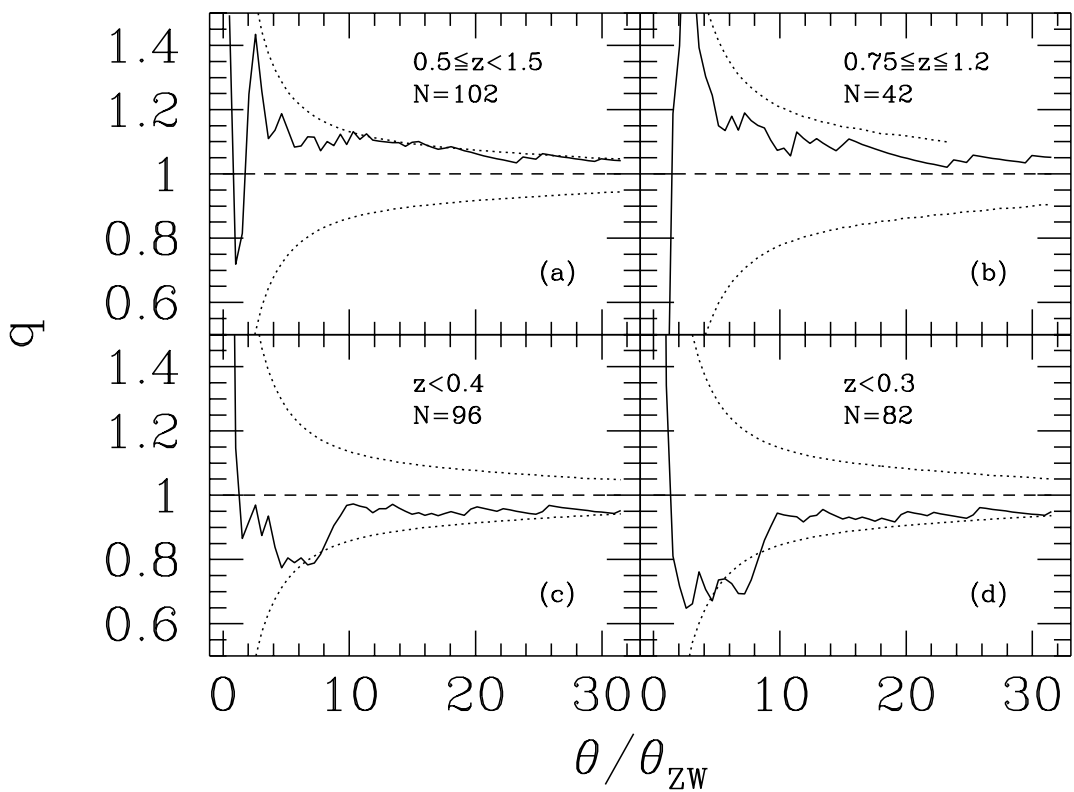

**Fig. 4.** Same as Fig. 3, but for different redshift intervals: (a): $0.5 \leq z \leq 1.5$ (102 quasars); (b): $0.75 \leq z \leq 1.2$ (42 quasars); (c): $z \leq 0.4$ (96 quasars); (d): $z \leq 0.3$ (82 quasars). Again, no optical flux threshold has been applied

objects relative to the foreground objects, given the total number of background objects in the sample. Therefore, we investigate the density enhancement in the following. In contrast to the overdensity, the density enhancement remains a meaningful quantity even for large values of $f$ for which the expected number $(1-f)N$ of quasars in the non-associated area becomes very small, and thus the corresponding density $\bar{n}_\mathrm{a}$ is subject to large statistical uncertainties. The overdensity and density enhancement are related to each other as follows:

$$\omega = 1 + \frac{q-1}{1-f\,q} \quad , \quad q = \frac{1}{f + (1-f)/\omega} \quad . \qquad (4)$$

In Figs. 3 to 9 we have plotted the density enhancements for the quasar sample and different subsamples; all curves are drawn on the same scales to allow for easy comparison. The solid curve in each panel displays the density enhancement $q$ as a function of $\theta/\theta_\mathrm{ZW}$, where $\theta = x\theta_\mathrm{ZW}$ defines the angular size of the cluster discs and thus the association area. We have calculated and plotted $q(x)$ for 61 equidistant values of $x$ in the interval $0.5 \leq x \leq 31.5$. To assess by eye what kinds of events are rare for a random distribution we have included two significance lines: for each value of $x = \theta/\theta_\mathrm{ZW}$ (or association area), the probability to find a density enhancement equal to or above (below) the upper (lower) dotted curve is less than or equal to 2 percent for a random point distribution. Hence, these 98% significance curves separate 'common' events from

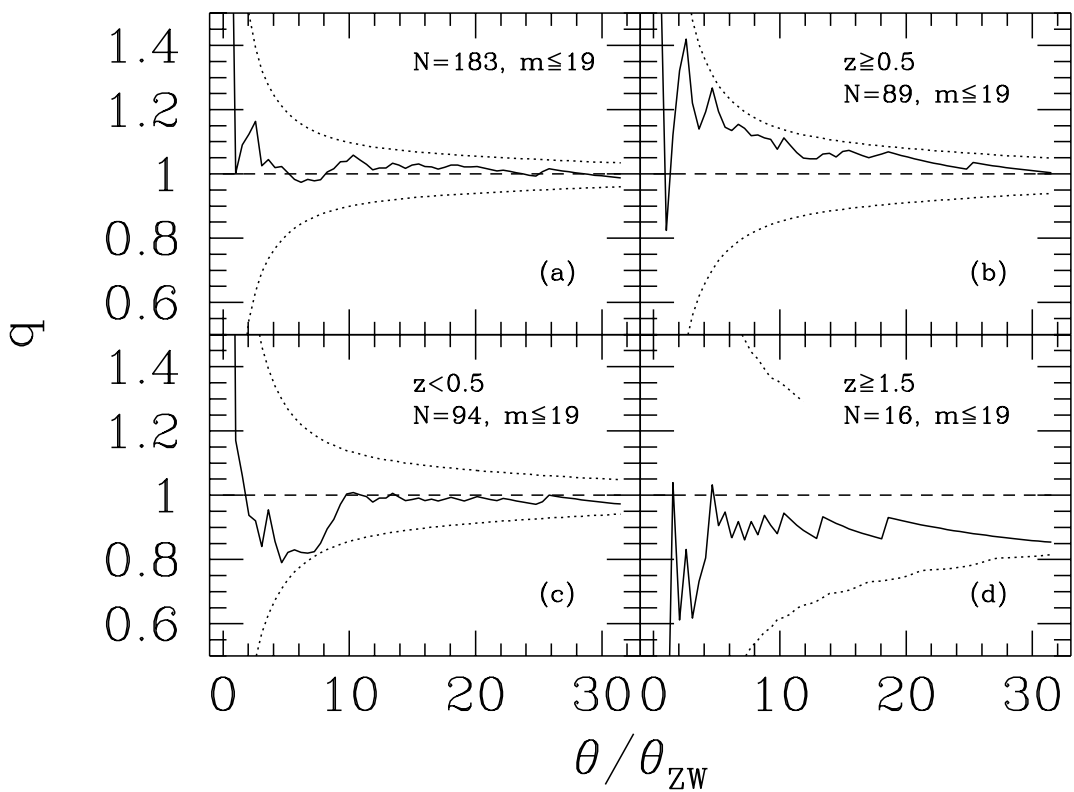

**Fig. 5.** Same as Fig. 3, but with an optical flux threshold of $m \leq 19$. The redshift intervals and number $N$ of quasars in each sample is written in each panel

'rare' ones. The dashed curve in each panel indicates the expectation value (1) of the density enhancement for a random distribution of points. We want to point out, that evaluating the probability for a given density enhancement, it is not adequate to use the Poisson distribution, since the expectation value of the number density of QSOs in the non-associated area is not known. Instead, what is known is the total number of QSOs; hence, the probability distribution which describes the number of QSOs within the associated area is the binomial distribution, for which the Poisson distribution in our case is not an adequate approximation, since (1) the number of QSOs is small and (2) the fractional association area $f$ is generally not small compared to unity. [1] The

---

[1]  This remark also applies for the RH94 analysis: On a scale of $\theta = 6\theta_{\mathrm{ZW}}$ the association area for their cluster sample covers 123 square degrees and contains $N_{\mathrm{a}} = 69$ QSOs, the remaining background area contains $N_{\bar{\mathrm{a}}} = 60$ QSOs in 181 square degrees; the fractional association area, the overdensity and the density enhancement are 0.4046, 1.7 and 1.3, respectively. The authors assume the expected quasar density in the non-associated area to be the observed one and evaluate from this the expectation value of QSO in the association area. The significance of the observed deviation from the expectation value is calculated in terms of the standard deviation $\sigma$, which is defined for an arbitrary probability distribution as the square root of the variance of this distribution; for a Poisson distribution the standard deviation simplifies to the square root of the expectation value. From this, RH94 estimate that counting 69 QSOs instead of 40.8 corresponds to a 4.7 $\sigma_{\mathrm{Poisson}}$ result. However, as noted before, the density of the non associated QSOs is also subject to statistical fluctuations, and therefore one can not deduce the expectation value for the QSO number in the association area from the QSO density in the non-association area. From the total number of QSOs (129), we expect 52.2 QSO on the association area, and, considering only the associated

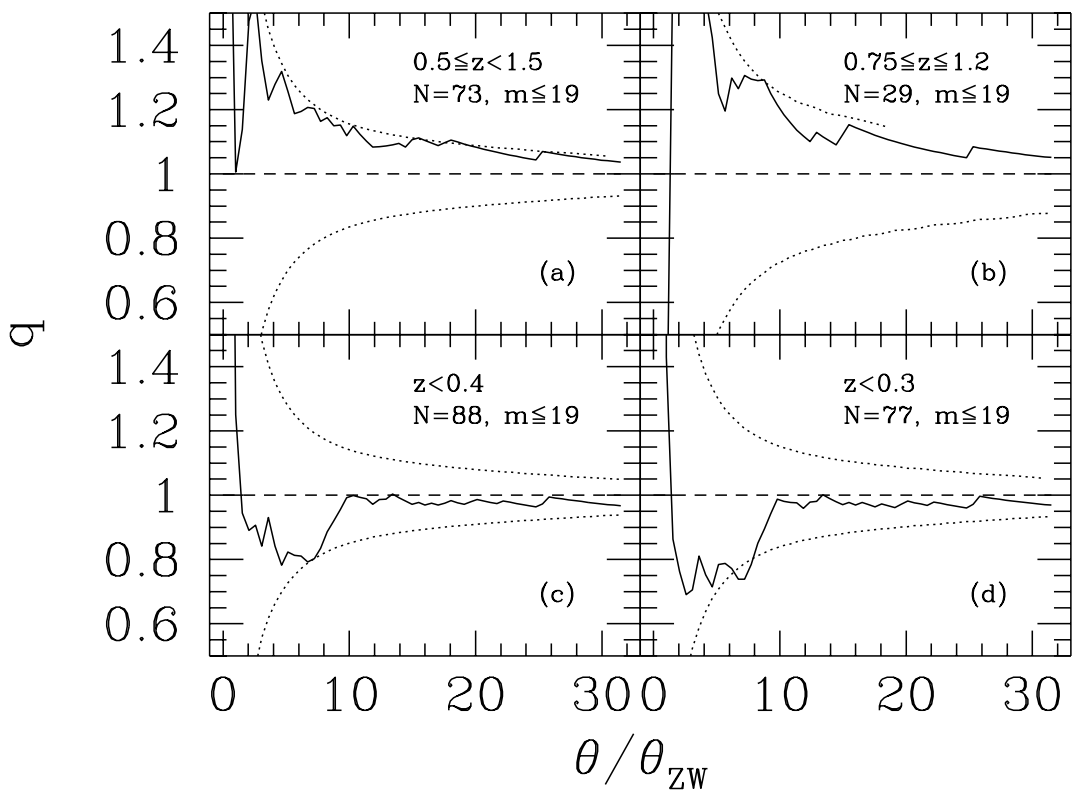

**Fig. 6.** Same as Fig. 4, but with an optical flux threshold of $m \leq 19$. The redshift intervals and number $N$ of quasars in each sample is written in each panel

binomial distribution yields the probability that one finds, given a fixed total number $N$ of quasars, $N_a(f)$ quasars in a fraction $f$ of the total area. Since this distribution is discrete, we obtained our smooth significance curves by a linear interpolation of the cumulative distributions $p(\leq n)$ and $p(\geq n)$. If the number $N$ of quasars in a sample is smaller than 78, the upper significance curve is no longer defined for the largest value of $x = \theta/\theta_{ZW}$ we considered, i.e., $x_{\max} = 31.5$, since the probability that all $N$ randomly positioned points are on the association area can then become larger than 2% (i.e., there is no density enhancement event with a probability of 2 percent or less). The smaller the quasar sample, the smaller the scale up to which the upper significance curve is defined (compare Figs. 8c,d, 7b, 8a, 4b, 6b, 3c, 8b, 5d and 7d where the corresponding quasar samples contain 73, 63, 54, 44, 42, 29, 25, 20, 16 and 10 quasars respectively. For the case of 10 quasars (Fig. 7d), the upper significance curve is only defined for $\theta/\theta_{ZW} \leq 8.77$ where its value is 1.5; therefore, this curve can not be seen in the displayed range of Fig. 7d.). Analogously, the lower significance curve is not defined for the smallest angular scales considered ($x_{\min} = 0.5$) if the number of quasars in the sample becomes

---

QSOs, the expectation value deviates from 69 QSOs by 2.3 $\sigma_{\text{Poisson}}$. However, the distribution describing the number of associated QSOs on a fraction $f$ of the total area and the number of non associated QSOs on the remaining area, is the binomial distribution; with 99.8 percent probability one finds fewer than 69 associated QSO (and more than 60 non associated QSOs). The standard deviation for a binomial distribution is $\sigma_{\text{binomial}} = \sqrt{N(1-f)f} = 5.3$ for a total number $N$ of QSOs. Hence, finding 69 of 129 QSO on 40 percent of the total area is a 3 $\sigma_{\text{binomial}}$ result.

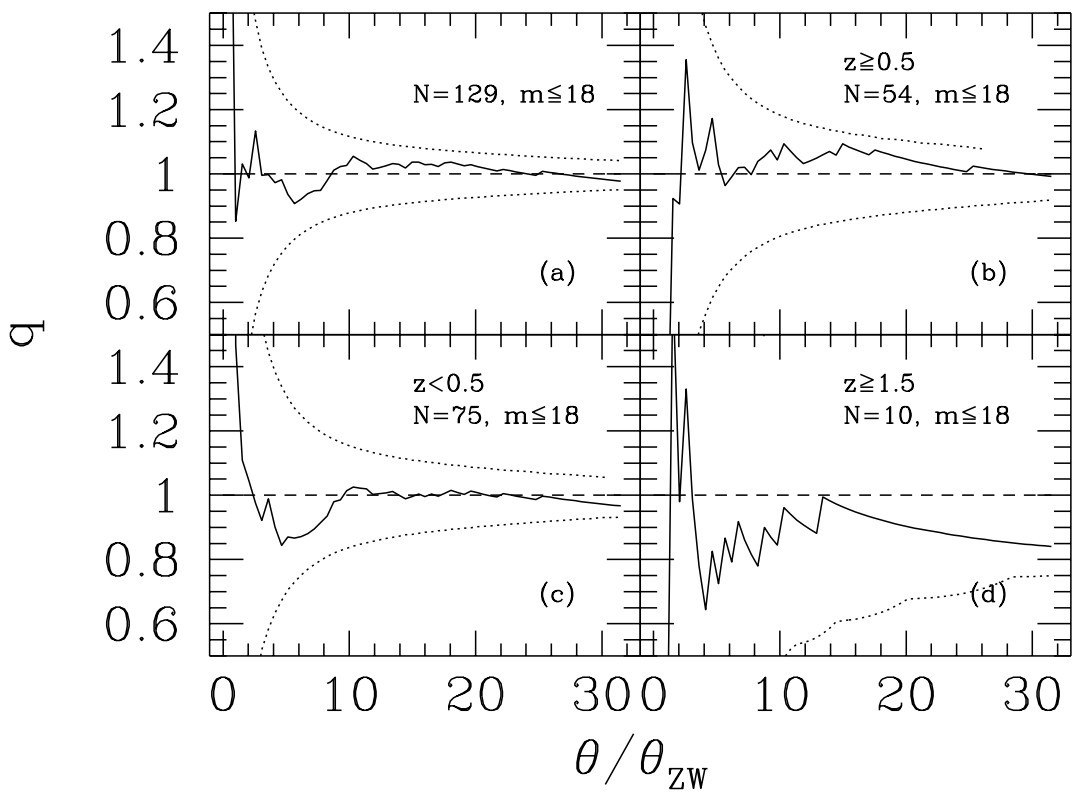

**Fig. 7.** Same as Fig. 3, but with an optical flux threshold of $m \leq 18$. The redshift intervals and number $N$ of quasars in each sample is denoted in each panel

smaller than 57. Then the probability that none of $N$ randomly positioned points is on the association area can become larger than 2 percent, i.e. there exists no density reduction event with a probability of 2 percent or less. This applies for the quasar sample shown in Fig.9, where the lower significance curve (i.e. the curve which indicates a significant density reduction) is defined only for $\theta/\theta_{\rm ZW} \gtrsim 3$. Of course, finding a scale where the enhancement curve is above (below) the upper (lower) significance curve does not imply that one has found a density enhancement (reduction) with a significance of more than 98 percent, since each curve consists of the results of 61 'trials' which are not mutually indepedent. We shall return to this point in detail later. Thus, these significance curves only intend to give a visual impression, and the significance of a given density enhancement curve defined on $0.5 \leq \theta/\theta_{\rm ZW} \leq 31.5$ is calculated by Monte Carlo simulations later on.

All enhancement curves show a broad scatter around unity on very small angular scales; this is no systematic effect as can be checked by considering the disjunct subsamples $[z \leq 0.5,\ z \geq 0.5]$ or $[z \geq 1.5,\ 0.5 \leq z \leq 1.5]$: the enhancement factor is either very high or very small compared to 1. Therefore, the scattering is statistical and is due to the small expectation value $Nf$ for associated quasars; this is also reflected in the strong increase and decrease of the upper and lower significance curves.

In Fig. 3a no strong deviation of the enhancement curve $q(\theta/\theta_{\rm ZW})$ from unity can be seen, whereas in the high (Fig. 3b) and low (Fig. 3c) redshift subsample there is a slight tendency for a density increase and decrease, respectively. The very high redshift sample ($z \geq 1.5$), Fig. 3d, is distributed like the median and therefore compatible to a random distribution. In none of the four quasar subsamples in Fig. 3 the 98% significance curves are touched or crossed. However, a prominent density enhancement arises for the remaining high redshift quasars: $0.5 \leq z \leq 1.5$, Fig. 4a. The density enhancement curve is above the 98% significance curve for 10 of the 61 grid points used: e.g., at $\theta/\theta_{\rm ZW} = 10.3$, expected. The enhancement factors are 1.13 and 1.125 (and translate to overdensities of 1.74 and 1.85). A further restriction of the quasars' redshift slice to $0.75 \leq z \leq 1.2$ yields a larger density enhancement (Fig. 4b), but at a reduced significance due to the small number (42) of quasars in this subsample. Now only at 1 of the 61 grid points (at $x = 2.5$) the enhancement curve is above the 98% significance curve.

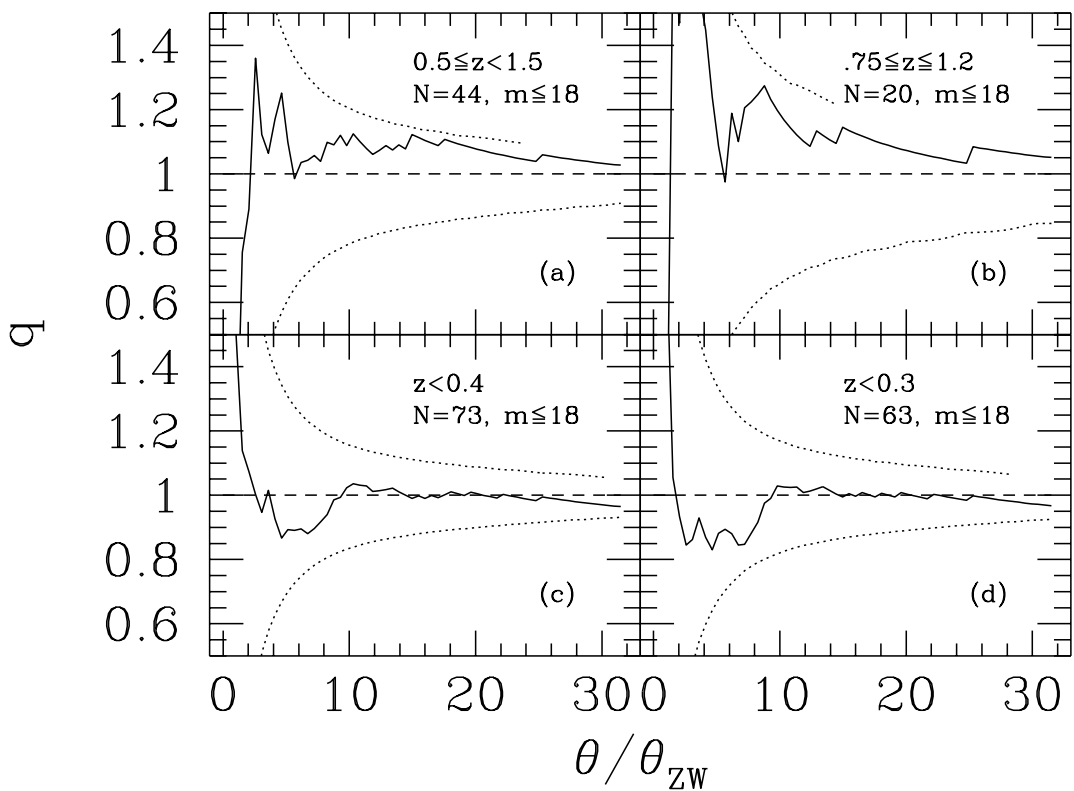

**Fig. 8.** Same as Fig. 4, but with an optical flux threshold of $m \leq 18$. The redshift intervals and number $N$ of quasars in each sample is written in each panel

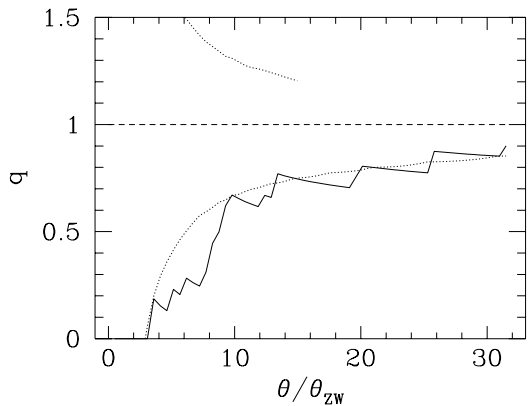

**Fig. 9.** Same as Fig. 3, but for a sample with $z \leq 0.3$, and optical flux fainter than $m \geq 18$, containing 21 quasars

Considering subsamples of the low redshift quasars, $z < 0.5$, $z < 0.4$ and $z < 0.3$, reveals an increasing deficiency of quasars around Zwicky clusters with lower redshift limit in size and significance, see Figs. 3c, 4c and 4d. For $z \leq 0.3$, the neighborhood ($\leq 8 \times \theta_{\rm ZW}$) of clusters shows a depletion of quasars with $z \leq 0.3$ quasars by about 30%. For $\theta/\theta_{\rm ZW} = 6.7, 7.2$ and $7.7$ we expect 44.7, 47.6 and 50.2 quasars but we find only 31, 33 and 37, respectively.

Up to now we have only considered quasar samples without an explicitly imposed optical flux threshold. As for the total quasar sample, the samples not restricted in redshift but limited to $19^{\rm th}$ and $18^{\rm th}$ optical magnitude (Figs. 5a and 7a) show no deviation from a uniform distribution. Due to the optical flux limit, the number of quasars decreases by 23% or 45%, respectively; therefore, a density enhancement for bright subsamples has to be larger to become significant. The redshift-selected, optically flux-limited samples show the same qualitative behaviour as the corresponding samples without an optical flux limit: high-redshift quasars with $z \geq 0.5$ are slightly but not significantly overabundant (Figs. 3b, 5b and 7b); the density enhancement is largest for the case of a flux limit of $19^{\rm th}$ magnitude. For brighter quasars, $m \leq 18$-th, the density enhancement weakens; in particular, quasars on small scales become less overabundant. As in the case of no optical flux limit, the very high redshift quasars with $z \geq 1.5$ are distributed like the median of a random distribution (Figs. 5b and 7b). Note that the very small quasar numbers in these subsamples (16 and 10 quasars) hardly allow for a statistically significant density enhancement or decrease. Since it is not appropriate in a statistical investigation to report only results on those samples which yielded a positive result in size and significance, we present the results on all quasar samples we have investigated, irrespective of the outcome of the statistical analysis.[2]

A flux limit at $19^{\rm th}$ magnitude for the redshift slice $0.5 \leq z \leq 1.5$ increases the density enhancement on scales $\theta \leq 10\theta_{\rm ZW}$ by $\approx 0.1$ relative to the case of no flux limit. There are 29 (48, 61) quasars on a scale of $\theta/\theta_{\rm ZW} = 4.1$ (6.7,10.3), whereas the expectation value is 22.7 (39.8, 53.1). 5 of the 61 gridpoints of the enhancement curve are on or above the 98% significance curve.

Although for $0.5 \leq z \leq 1.5$, the enhancement curves are similar for the case of no flux limit and for $m \leq 18$, the latter never crosses the 98% significance curve, since this sample contains only half the number of quasars of the first one.

For the quasars at $0.75 \leq z \leq 1.2$, the enhancement increases with a flux limit at $19^{\rm th}$ magnitude in size by about 0.1. However, the upper significance curve is only defined up to $\theta/\theta_{\rm ZW} = 18.6$ and thus the density enhancement curve exceeds the significance curve objects ($m \leq 18$), but is still larger than for sources without a flux limit; the 98% significance curve is crossed nowhere.

---

[2] We want to point out that the analysis of RH94 is restricted to a single scale, corresponding to $x = 6$; however, this particular scale was selected by considering the angular two-point correlation function (in units of the $\theta_{\rm ZW}$s between their QSOs and the Zwicky clusters). Hence, in effect, they have tested more than one angular scale with the angular correlation method from which they selected the one with the largest signal for their subsequent investigation for the overdensity. The meaning of the statistical significance of their result is therefore not easy to interpret.

The density reduction for the low redshift sample $z \leq 0.5$ continuously decreases for optically brighter quasars, see Figs. 3c, 5c and 7c. This also applies for a lowered redshift limit of 0.4 and 0.3, as can be seen in Figs. 4c,d, 5c,d and 6c,d. Faint low-redshift radio sources, $z \leq 0.3$ , $m \geq 18$, show a large deficiency of about 80% for $\theta \leq 8\theta_{\rm ZW}$.

After the qualitative inspection of the enhancement curves (by eye), eight subsamples show a strong deviation from a random distribution: quasars in the redshift intervals $0.5 \leq z \leq 1.5$ (with no flux limit, and with $m \leq 19$), $0.75 \leq z \leq 1.2$ (with no flux limit, and with $m \leq 19$) are overdense behind Zwicky clusters; radio sources with $z \leq 0.3$ (with no flux limit, and with $m \leq 19$), with $z < 0.4$ (no flux limit), and with $z < 0.3$ and $m \geq 18$ are underdense close to the line of sight to Zwicky clusters. We have calculated the significance for these density deviations for each individual subsample. Considering an enhancement curve one must not pick a scale where the density enhancement is especially high and calculate the probability of the corresponding enhancement on that particular scale! Instead one should evaluate the probability to obtain a certain enhancement curve, i.e., one also includes the results on those scales where the enhancement is not particularly high. To do this, one has to take into account that neighboring points of an enhancement curve are not statistically independent. Hence we obtained our significance levels by Monte Carlo simulations: for a subsample with $N$ quasars and an enhancement curve which is on or above (on or below) the upper (lower) 98% significance curve at $j$ of the 61 grid points, we have calculated the probability, $P_N(\leq j-1)$ that a random distribution with $N$ quasars falls on or above (on or below) this curve at $j-1$ or fewer gridpoints, by using $M$ realizations of $N$ randomly positioned points; typically, $M = 1000$, but for the two most significant subsamples in Tables 1 & 2, we took $M = 10000$. We denote this probability as the significance of an observed result. We present our results in tables 1 and 2:

**Table 1.** For four different quasar samples, as described in the first column, we show the number of gridpoints $x_i$ at which the density enhancement curve lies on or above the upper significance curve (second column), and the corresponding significance level as determined from Monte-Carlo simulations – see text

| Sample parameters | NR of gridpoints on/above the upper signficance curve | Significance for for a density enhancement |
|---|---|---|
| $0.5 \leq z < 1.5$ , no flux limit | 10 | $P_{102}(j \leq 9) = 97.67$ |
| $0.75 \leq z < 1.2$ , no flux limit | 1 | $P_{42}(j \leq 0) = 89.1$ |
| $0.5 \leq z < 1.5$ , $m \leq 19$ | 5 | $P_{73}(j \leq 4) = 95.9$ |
| $0.75 \leq z < 1.2$ , $m \leq 19$ | 1 | $P_{29}(j \leq 0) = 87.8$ |

From all four samples which show a density enhancement, that with the largest quasar number yields the highest significance (97.67 percent); due to the decrease of quasar numbers, the significance for a density enhancement decreases if one imposes a flux limit or reduces the redshift interval, although the enhancement curves are similar or even higher. The small number of quasars does not allow to further subdivide the quasar samples and old. This is not true for the quasar samples which are underdense near Zwicky clusters: Decreasing the upper cutoff in redshift strenghtens the density reduction in amount and significance. This means that the very low redshift quasars $z < 0.3$ are most responsible for the lack of quasars in the direction of clusters. Since an optical flux limit weakens the density reduction, we also investigated the distribution of the optically faint, $m \geq 18$, low-redshift quasars at $z < 0.3$ : these are depleted in the direction of Zwicky clusters with a significance above 99.89 percent.

Summarizing our results, we have found that low-redshift ($z < 0.4$) quasars are underdense near the line of sight to Zwicky clusters, that high-redshift quasars ($0.5 \leq z \leq 1.5$) are overdense behind Zwicky clusters and that very high-redshift quasars are uniformly distributed with respect to the clusters.

**Table 2.** Same as Table 1, for four other samples; in this case, the second column shows the number of gridpoints $x_i$ at which the density enhancement curve lie on or below the lower significance curve

| Sample parameters | NR of gridpoints on/below the lower signficance curve | Significance for for a density reduction |
|---|---|---|
| $z < 0.3$ , no flux limit | 7 | $P_{82}(j \leq 6) = 96.3$ |
| $z < 0.3$ , $m \leq 19$ | 3 | $P_{77}(j \leq 2) = 93.3$ |
| $z < 0.4$ , no flux limit | 3 | $P_{96}(j \leq 2) = 93.7$ |
| $z < 0.3$ , $m \geq 18$ | 39 | $P_{21}(j \leq 38) = 99.89$ |

## 5 Interpretation

Correlations of non-physically associated objects are most frequently explained by selection effects, obscuration by dust, or by gravitational lensing. We now have to find mechanisms which can account for an anticorrelation of low-redshift quasars, a correlation of high-redshift quasars, and no correlation of very high-redshift quasars with respect to Zwicky clusters.

A positive correlation of two optically selected populations (QSO and clusters) can be explained by patchy dust in our Galaxy: the number counts of both populations are depleted in the direction of dust patches and are unaffected in the remaining directions; i.e. one either observes a high or a low number density of *both* populations and quantifies this by a correlation strength or in terms of a density enhancement. The number counts for radio selected objects, like those in the 1-Jansky catalog, do not suffer from dust obscuration. However, dust renders radio source identifications and redshift measurements more difficult. Therefore, if there is patchy dust in our Galaxy, radio selected objects with determined redshift will show a small positive correlation to optically selected clusters. This correlation will become stronger (and will be equal to that of optically selected objects) if an additional optical flux limit is imposed on the radio-selected objects. This yields the expectation that, *independent of redshift*, the quasars are weakly (for the case of no the Zwicky clusters. It is clear that this scenario fails to explain the anticorrelation of low redshift quasars and the uniform distribution of very high redshift quasars.

Anticorrelation of emission-line selected QSO's to 'clusters' (Boyle et al. 1983, 1984) can be explained by the optically crowded fields, which makes the inspection of objective prism plates difficult. For QSOs selected by a different optical criterion this interpretation dose not apply, instead, dust in clusters is made responsible for the underdensity, (see Romani & Maoz 1992, Boyle et al. 1988). Dust in clusters makes all objects lying behind or in the clusters underdense with respect to the densities of objects not lying in the direction of clusters. As before, this is also valid for radio-selected objects with an additional optical flux limit, and, in a weaker sense, still for radio objects with measured redshift. Hence, dust in Zwicky clusters (with a mean redshift of 0.2) *depletes all quasars with $z \gtrsim 0.2$, to a degree which is independent of z*. Hence, it can not explain the positive correlation of high-redshift quasars, and the uniform distribution of the very high-redshift quasars. Also, dust in our Galaxy and dust in clusters can not conspire to explain the observations.

Lensing provides the only known szenario which can qualitatively explain the overdensity of high-redshift quasars behind Zwicky clusters *and* the uniform distribution of very high-redshift quasars with $z \geq 1.5$. Density inhomogeneities between a source and an observer enhance the source flux; thus objects (QSOs) which are too faint in the absence of lenses can be magnified above the flux threshold and end up in a flux-limited QSO catalog ('magnification bias'). A crude but useful quantification for the relation of the unlensed $n_{\mathrm{unlensed}}(> S)$ and the lensed number counts of QSOs $n_{\mathrm{lensed}}(> S)$ above a flux $S$ is

$$n_{\mathrm{lensed}}(> S) \gtrsim \mu_{\mathrm{mean}}^{\alpha-1} \, n_{\mathrm{unlensed}}(> S) \quad , \tag{5}$$

where $\alpha$ is the absolute value of the double-log slope of the integrated number counts of the source population and $\mu_{\mathrm{mean}}$ is the average magnification caused by the intervening mass. For galaxies counted in the optical, $\alpha$ is equal to one, for optically selected QSOs it is of the order of 2.6 (for bright QSOs); 1-Jansky source counts with an optical flux limit of $19^{\mathrm{th}}$ ($18^{\mathrm{th}}$) magnitudes have an effective slope $\alpha$ of 3 (3.5) (see Bartelmann 1994) if the multiple-waveband magnification bias is employed (Borgeest, von Linde & Refsdal 1991). The mean magnification (or the lensing strength) of a density inhomogeneity does not only depend on its mass distribution but is also proportional to $D_{\mathrm{d}} D_{\mathrm{ds}} / D_{\mathrm{s}}$, i.e. to the product of the angular diameter distances from the observer to the lens, the lens to the source, divided by the angular diameter distance from the observer to the source. For a fixed source redshift, lenses in a redshift range where $D_{\mathrm{d}} D_{\mathrm{ds}}$ is maximal are most efficient. Mass distributions at $z \approx 0.2$ are the most efficient lenses for QSOs at a redshift of slightly smaller than one, whereas masses at 0.4 to 0.5 most efficiently magnify QSOs with $z \approx 1.5$ to 2. The clusters in the Zwicky catalog trace inhomogeneities with a maximum redshift of about at most 0.3; hence we expect quasars with a redshift of not more than one to be magnified by foreground Zwicky clusters. Quasars with very high redshifts may also be overdense near to the line of sight to $z \approx 0.5$ lenses, but these lenses are not visible on the plate material of the Palomar Sky Survey. In other words, for very high-redshift quasars the most effective lenses decouple from visible lenses in the Zwicky catalog and therefore we expect very high-redshift quasars to be nearly randomly distributed with respect to Zwicky clusters.

$19^{th}$ mag. relative to the case of no flux limit can also be explained by gravitational lensing. Since the number counts of 1-Jansky sources with optical flux threshold are effectively steeper than without a flux limit, we expect the magnification bias to be more efficient in the case of a brighter quasar sample. We have seen in Sect. 4 that the significance of the density enhancement for the brighter quasar samples weakens due to the smaller number of quasars in the sample; therefore we only want to point out that we can at least interpret the tendency that brighter samples show a larger density enhancement than fainter ones, by gravitational lensing. The most important point is, however, that gravitational lensing explains the redshift dependence of the correlations observed (see also Sect. 6).

Low-redshift 1-Jy radio sources are depleted near Zwicky clusters. Most of these sources are actually radio galaxies. In this case, lensing as an explanation can be safely excluded. As we have argued before, this depletion is highly unlikely to be due to dust obscuration: if patchy dust were in our Galaxy, obscuration would rather yield a positive correlation, whereas if the obscuration occured in the clusters, it would also affect high-redshift quasars, in contrast to our finding (of course, one could argue that by a delicate balance of lensing and obscuration, the observed redshift dependence can be explained, but we consider this possibility to be unnatural; in addition, in this case the lensing effect needs to be much larger). Since the largest effect comes from sources with $z \leq 0.3$, those are most likely at the same distance as the clusters themselves. The observed underdensity can thus be most easily interpreted as a physical effect which signifies the influence on the environment of an AGN on its activity. Hence, our finding would imply that radio galaxies tend to avoid high galaxy densities. This result, however, is in contrast with the fact that low-redshift QSOs tend to prefer higher density environments (Ellingson, Yee & Green 1991). Since we find no easy explanation for this apparent discrepancy, we will end this discussion here.

# 6 Discussion

In this paper we have investigated the correlation between distant Zwicky clusters and 1-Jy radio sources. We find a statistically significant overdensity of quasars with $z \approx 1$ around Zwicky clusters; higher-redshift quasars seem to be distributed randomly with respect to the cluster sample. This redshift behaviour is in complete agreement with the finding of Bartelmann & Schneider (1993) that 1-Jy quasars with $z \approx 1$ are associated with Lick galaxies, and that the correlation decreases for higher redshifts. This agreement is reassuring, since perhaps the weakest point of both analyses are the selection criteria for the clusters (or galaxies). In both cases this has been done by eye inspection and it is thus affected by subjective selection. However, the fact that the Lick sample has been defined independently of the current cluster sample means that these selection effects probably do not dominate our results. Our results are at variance with those of RH94, who obtained the statistically significant overdensity for QSOs at redshift $1.4 \leq z \leq 2.2$. However, their QSO sample has been selected in a completely different way; in particular, they used optically-selected sources, which are more affected by possible dust obscuration and crowded fields. This latter point might indicate a partial explanation for their finding have discussed in this paper, the significance of the RH94 result may be too optimistic.

Since the 1-Jy catalog is not completely identified yet, and 'only' about 83% of its sources have their redshift measured, one must consider the possibility that the selection of those sources for which the redshift is measured (and which we thus have used in our analysis) affects the statistical results obtained here. However, the completeness of the identification and redshift determination of the 1-Jy sample is much larger in the northern hemisphere, and most of the sources there without redshift are either empty field sources, or faint radio galaxies. It is therefore highly unlikely that the selection of sources with determined redshift affects the statistical results derived here (H. Kühr, private communication).

The redshift dependence of the correlations for the high-redshift quasars is understood qualitatively within the gravitational lensing hypothesis (see also Bartelmann & Schneider 1993a,b). However, since the effects we consider occur on angular scales of order one degree, a back-of-the-envelope estimate on the effectiveness of (isothermal) cluster lenses does not explain the observed strength of the correlations (see laso RH94). However, we believe that the true situation is much more complicated: on the angular scales on which the correlations are seen the Zwicky cluster heavily overlap. Furthermore, the lensing effects investigted in Bartelmann & Schneider (1993a) were mainly caused by the weak magnifications by density inhomogeneities on scales larger than that of clusters, so that the clusters need to account only for part of the lensing effect; they are further assisted by the larger-scale matter distributions.

For low-redshift radio sources, we found a highly significant underdensity around Zwicky clusters. Since most of these sources are radio galaxies at redshifts not much smaller than the mean redshift of our cluster sample, we suspect that this underdensity can be interpreted as an effect of the cluster environment on the nuclear activity in those sources, in the sense that at such redshifts, high-luminosity radio galaxies avoid a cluster environment.

**Acknowledgments**

We thank Matthias Bartelmann for helpful discussions and for constructive comments on the manuscript.